\documentclass{article}
\usepackage{times}
\usepackage{amsfonts}
\usepackage{graphicx}
\usepackage[pdfmark]{hyperref}
\begin{document}
\noindent
{\Large  DIRICHLET BRANES AND A COHOMOLOGICAL\\ DEFINITION OF TIME FLOW}
\vskip1cm
\noindent
{\bf Jos\'e M. Isidro and P. Fern\'andez de C\'ordoba}\\
Grupo de Modelizaci\'on Interdisciplinar Intertech,\\
 Instituto Universitario de Matem\'atica Pura y Aplicada,\\ Universidad Polit\'ecnica de Valencia, Valencia 46022, Spain\\
{\tt joissan@mat.upv.es, pfernandez@mat.upv.es}
\vskip1cm
\noindent
\today
\vskip1cm
\noindent
{\bf Abstract} Dirichlet branes are objects whose transverse coordinates in space are matrix--valued functions. This leads to considering a matrix algebra or, more generally, a Lie algebra, as the classical phase space of a certain dynamics where the multiplication of coordinates, being given by matrix multiplication, is nonabelian. Further quantising this dynamics by means of a $\star$--product introduces noncommutativity (besides nonabelianity) as a quantum $\hbar$--deformation. The algebra of functions on a standard Poisson manifold is replaced with the universal enveloping algebra of the given Lie algebra. We define generalised Poisson brackets on this universal enveloping algebra, examine their properties, and conclude that they provide a natural framework for dynamical setups (such as coincident Dirichlet branes) where coordinates are matrix--valued, rather than number--valued, functions.

\vskip1cm
\section{Introduction}\label{labastidachupamelapolla}

Classical mechanics can be formulated on a Poisson manifold ${\cal M}$ (classical phase space). This means that the algebra $C^{\infty}({\cal M})$ of smooth functions on ${\cal M}$ supports Poisson brackets, {\it i.e.}, an antisymmetric, bilinear map
\begin{equation}
\left\{\cdot\,,\cdot\right\}_{\rm Poisson}\colon C^{\infty}({\cal M})\times C^{\infty}({\cal M})\longrightarrow C^{\infty}({\cal M})
\label{labastidaquetepartaunrayo}
\end{equation}
satisfying the Jacobi identity and the Leibniz derivation rule \cite{RATIU}.

On the other hand, Lie algebras $\mathfrak{g}$ support Lie brackets
\begin{equation}
[\cdot\,,\cdot]\colon \mathfrak{g}\times \mathfrak{g}\longrightarrow \mathfrak{g}
\label{ramallocabron}
\end{equation}
that are antisymmetric, bilinear and satisfy the Jacobi identity. For reasons that will become clear presently, we would like to regard $\mathfrak{g}$ as a classical phase space for a certain dynamics, and its universal enveloping algebra ${\cal U}(\mathfrak{g})$ as its {\it algebra of functions}, in a sense to be specified below. 

When $\mathfrak{g}$ is simple, we can use the Killing metric to identify $\mathfrak{g}$ with its dual $\mathfrak{g}^*$. 
Then the Kirillov brackets on $\mathfrak{g}^*$,
\begin{equation}
\left\{\cdot\,,\cdot\right\}_{\rm Kirillov}\colon C^{\infty}(\mathfrak{g}^*)\times C^{\infty}(\mathfrak{g}^*)\longrightarrow C^{\infty}(\mathfrak{g}^*)
\label{barbonchupamelapolla}
\end{equation}
turn $\mathfrak{g}$ into a Poisson manifold \cite{KIRILLOV}. Although we will make use of them, we are not primarily interested in the Kirillov brackets (\ref{barbonchupamelapolla}). Instead we will pass from the Lie algebra $\mathfrak{g}$ to its universal enveloping algebra ${\cal U}(\mathfrak{g})$, where an associative multiplication is defined, and the Leibniz derivation rule can be made to hold. The price to pay is the loss of the abelian property for the associative multiplication: while the algebra $C^{\infty}({\cal M})$ was abelian under the pointwise multiplication of functions, ${\cal U}(\mathfrak{g})$ will not be abelian. This reflects the nonabelian property of $\mathfrak{g}$, which in turn is necessary for $\mathfrak{g}$ to be simple.

Evolution equations of the type
\begin{equation}  
\dot F = \left[F, H\right],
\label{labastidaputonverbenero}
\end{equation}
for $F\in\mathfrak{g}$ and for a certain Hamiltonian $H\in\mathfrak{g}$, are ubiquitous in physics. Interesting generalisations of the above equation can be obtained as follows. For definiteness we will consider the Lie algebra $\mathfrak{su}(n)$ throughout, although our results can be easily generalised to any finite--dimensional, simple, compact Lie algebra $\mathfrak{g}$, provided one pays due attention to its corresponding cohomology ring \cite{KNAPP}. (Since $\mathfrak{g}$ is compact we may alternatively consider de Rham cohomology on the corresponding Lie group \cite{GOLDBERG}). Let $\omega_{2j+1}$ be a nonzero cocycle of the cohomology of $\mathfrak{g}$, with order $2j+1$. This cocycle defines  $2j$--fold Lie brackets on $\mathfrak{g}^{\times (2j)}$,
\begin{equation}
[\cdot\,,\cdots,\cdot]_{\omega_{2j+1}}\colon \mathfrak{g}\times{}^{(2j)\atop\ldots}\times \mathfrak{g}\longrightarrow
\mathfrak{g},
\label{pajarescabron}
\end{equation}
linear in all $2j$ entries, completely antisymmetric, and satisfying a generalised Jacobi identity \cite{BUENO}.
When $j=1$, the $2j$--fold Lie brackets (\ref{pajarescabron}) reduce to the Lie brackets (\ref{ramallocabron}), where one omits the subindex $\omega_3$. Then an equation of motion for $F\in\mathfrak{g}$ in the dynamics generated by $2j-1$ Hamiltonians $H_2, \ldots H_{2j}\in\mathfrak{g}$ is
\begin{equation}
\dot F = \left[F,H_2, \ldots, H_{2j}\right]_{\omega_{2j+1}}.
\label{sanchezguillencabron}
\end{equation}
We will generalise eqns. (\ref{labastidaputonverbenero}) and (\ref{sanchezguillencabron}) by allowing $F$ and the Hamiltonians to be elements of ${\cal U}(\mathfrak{g})$. Such evolution equations allow one to regard $\mathfrak{g}$ as a classical phase space, and the universal enveloping algebra ${\cal U}(\mathfrak{g})$ as an algebra of functions. We will define  $2j$--fold Poisson brackets on ${\cal U}(\mathfrak{g})$ that will satisfy the Leibniz derivation rule, and that will reduce to the $2j$--fold Lie brackets (\ref{pajarescabron}) when acting on elements of $\mathfrak{g}$. With this formal viewpoint, the impossibility of setting ${\cal U}(\mathfrak{g})$ (for $\mathfrak{g}$ compact and simple) equal to $C^{\infty}({\cal M})$ for any smooth Poisson manifold ${\cal M}$ becomes irrelevant. Of course, the restriction to the Cartan subalgebra $\mathfrak{h}\subset\mathfrak{g}$ leads to the abelian ${\cal U}(\mathfrak{h})$, a subalgebra of the full enveloping algebra ${\cal U}(\mathfrak{g})$; if $\mathfrak{g}$ has rank $l$, then ${\cal U}(\mathfrak{h})$ is the subalgebra of polynomials within $C^{\infty}(\mathbb{R}^l)$. 

In our terminology, {\it abelian}\/ and {\it commutative}\/ are not interchangeable, nor are their opposites {\it nonabelian, noncomutative}. We reserve the term {\it noncommutative}\/ for those multiplications performed using the $\star$--product \cite{GRONEWALD, MOYAL, BAYEN, KONTSEVICH, GMS}; {\it commutative}\/ are those products that use the pointwise multiplication of functions. On the other hand, matrix multiplication is termed {\it nonabelian}, although it remains commutative because it uses the pointwise product. A quantum deformation of matrix multiplication, eqn. (\ref{labamaricon}), will yield a multiplication that will be both nonabelian (because we will be dealing with matrices) and {\it noncommutative} (because of the $\hbar$--deformation). To illustrate our terminology, the algebra of functions $C^{\infty}({\cal M})$ is abelian and commutative. Replacing the pointwise product with a $\star$--product we obtain $C^{\star}({\cal M})$, which  is abelian and noncommutative. The universal enveloping algebra ${\cal U}(\mathfrak{g})$ is nonabelian and commutative; its quantum deformation ${\cal U}^{\star}(\mathfrak{g})$  will be nonabelian and noncommutative. All these algebras are associative.

We will address the deformation quantisation of the previous classical dynamics, by replacing the (pointwise)  multiplication on ${\cal U}(\mathfrak{g})$ with a Kontsevich $\star$--product. The latter requires the specification of  classical Poisson brackets. Roughly speaking, there is a 1--to--1 correspondence between classical Poisson structures and quantum $\star$--products \cite{KONTSEVICH}, hence the quantisation obtained depends on the classical brackets one starts out with. This quantisation can be carried out for any compact, simple Lie algebra $\mathfrak{g}$, but it is best performed by specifying a faithful representation for $\mathfrak{g}$ and regarding the matrices so obtained as {\it nonabelian coordinate functions}, their multiplication being nonabelian and, after quantisation, also noncommutative. One is thus led to the conclusion that $n$ coincident, parallel Dirichlet branes (D--branes for short) \cite{SZABO}, having matrices as their {\it transverse}\/ coordinates \cite{WITTEN}, provide a natural realisation of the abstract setup described previously. Moreover, the identification established in ref. \cite{WITTEN} between transverse components of bulk gauge fields and D--brane coordinates suggests the adjoint representation as the preferred one, but our treatment holds in any faithful representation just as well. 

Generalisations of the standard Poisson brackets involving more than 2 entries, in particular the Nambu brackets \cite{NAMBU}, have appeared in connection with branes and integrable systems \cite{CZ, HOPPE}. The approach of ref. \cite{CZ} is based on the observation that the worldvolume element on a membrane, being a Jacobian determinant, can be identified with the Nambu brackets. Although we also quantise by replacing the pointwise product with a $\star$--product, our approach differs from that of ref. \cite{CZ} in several respects. We do not regard the longitudinal brane coordinates are the basic variables entering the brackets. Instead our starting point is motivated in the consideration of matrix--valued functions as {\it transverse}\/ coordinate functions to the brane. Thus our approach is strongly motivated in M--theory \cite{SZABO}, where the Lie algebra in which coordinates take values plays a prominent role: it is the Lie algebra of the gauge group within the stack of coincident D--branes \cite{WITTEN}. In turn, the gauge symmetry $\mathfrak{su}(n)$ is determined only by the brane content, {\it i.e.}, by the number $n$ of coincident branes. (One may eventually add orientifolds in order to obtain an orthogonal/symplectic gauge symmetry within the branes, but we will basically consider the case of $\mathfrak{su}(n)$). 
In other words, transverse D--brane coordinate functions are determined by Yang--Mills gauge fields within the D--branes themselves. Moreover, being Lie--algebra valued, our transverse coordinate functions exhibit nonabelianity already at the classical level. 

Since we are addressing the mechanics (classical and quantum) of {\it transverse}\/ coordinates, and time is always {\it longitudinal}, or parallel to a brane, we can think of our construction as providing the mechanics (classical and quantum) of {\it matrix--valued, spacelike coordinate functions}\/ and their time evolution---in the absence of time! Indeed we will see that time evolution can be defined algebraically, by means of Lie algebra cohomology, without any recourse to a continuously flowing parameter.

This article is organised as follows. Section \ref{casposoramallo} presents the mathematical prerequisites concerning (classical and quantum) Lie and Poisson multibrackets \cite{BUENO}. Section \ref{labastidaquetefollen} works out the connection between multibrackets and $n$ parallel D$p$--branes (coincident or separated), where the gauge symmetry is $\mathfrak{u}(n)$ or a simple subalgebra thereof. Section \ref{mierdaparatodos} presents our conclusions.

\section{Dynamics on a simple, compact Lie algebra $\mathfrak{g}$}\label{casposoramallo}

Let $\mathfrak{g}$ be a simple, finite--dimensional, compact Lie algebra over $\mathbb{R}$; as a rule we have $\mathfrak{su}(n)$ in mind.

\subsection{Classical}\label{barbonputon}

The universal enveloping algebra ${\cal U}(\mathfrak{g})$ is the associative algebra obtained as the $\mathbb{R}$--linear span of all formal products of powers $X^{p}$, for all $X\in \mathfrak{g}$ and all $p=1,2,\ldots$, subject to the requirement that
\begin{equation}
XY-YX=[X,Y] \qquad \forall X, Y \in \mathfrak{g}.
\label{cesargomezmekagoentuputakaramarikondemierda}
\end{equation}
In eqn. (\ref{cesargomezmekagoentuputakaramarikondemierda}), the left--hand side contains the associative product on ${\cal U}(\mathfrak{g})$, while the the right--hand side contains the Lie brackets on $\mathfrak{g}$. 

The Lie brackets (\ref{ramallocabron}) extend to Poisson brackets 
\begin{equation}
\{\cdot\,,\cdot\}\,\colon{\cal U}(\mathfrak{g})\times{\cal
U}(\mathfrak{g})\longrightarrow {\cal U}(\mathfrak{g})
\label{ramallocasposo}
\end{equation}
by setting
\begin{equation}
\left\{X,Y\right\}:=\left[X,Y\right]\qquad \forall X,Y\in\mathfrak{g},
\label{barbonmaricondeplaya}
\end{equation}
by requiring multilinearity, complete antisymmetry, and by imposing the Leibniz derivation rule when applied to products of Lie algebra elements, {\it i.e.},
\begin{equation}
\left\{XY, Z\right\}:=\left\{X, Z\right\}\,Y + X\, \left\{Y, Z\right\}\qquad \forall X,Y,Z\in\mathfrak{g}.
\label{labastidacasposo}
\end{equation}
Higher powers of Lie algebra elements can be reduced to smaller powers by repeated application of eqn. (\ref{labastidacasposo}).
Picking a Hamiltonian $H\in{\cal U}(\mathfrak{g})$, the evolution equation for any $F\in{\cal U}(\mathfrak{g})$, or classical equation of motion, reads
\begin{equation}
\dot F= \left\{F, H\right\}.
\label{ramalloquetepartaunrayo}
\end{equation}

Choose now a $(2j+1)$--cocycle $\omega_{2j+1}$ in the cohomology ring of $\mathfrak{g}$. Then the $2j$--fold Lie brackets (\ref{pajarescabron}) can be extended to $2j$--fold Poisson brackets on ${\cal U}(\mathfrak{g})^{\times(2j)}$,
\begin{equation}
\{\cdot\,, \cdots,\cdot \}_{\omega_{2j+1}}\colon {\cal U}(\mathfrak{g})\times{}^{(2j)\atop\ldots}\times {\cal U}(\mathfrak{g})
\longrightarrow {\cal U}(\mathfrak{g}),
\label{pajareshijodeputa}
\end{equation}
by setting
\begin{equation}
\left\{X_1, \ldots, X_{2j}\right\}_{\omega_{2j+1}}:=\left[X_1,\ldots, X_{2j}\right]_{\omega_{2j+1}}
\label{barbonladron}
\end{equation}
for all $X_1, \ldots, X_{2j}\in \mathfrak{g}$, by demanding $2j$--linearity, complete antisymmetry, and 
further requiring that the Leibniz derivation rule hold. If we pick $2j-1$ Hamiltonians $H_2, \ldots, H_{2j}\in{\cal U}(\mathfrak{g})$ we can write a classical equation of motion for any $F\in {\cal U}(\mathfrak{g})$:
\begin{equation}
\dot F = \left\{F, H_2,\ldots, H_{2j}\right\}_{\omega_{2j+1}}.
\label{ramayoquetepartaunrayo}
\end{equation}
The $2j$--fold Poisson brackets (\ref{pajareshijodeputa}) reduce to the $2j$--fold Lie brackets (\ref{pajarescabron}) when restricted to $\mathfrak{g}^{\times (2j)}$, and to the Poisson brackets (\ref{ramallocasposo}) when $j=1$. In this latter case one omits the subindex $\omega_3$. 

In this way the algebra ${\cal U}(\mathfrak{g})$ supports the Poisson multibrackets (\ref{pajareshijodeputa}) even if ${\cal U}(\mathfrak{g})$ cannot be identified with the algebra of smooth functions $C^{\infty}({\cal M})$ for any manifold ${\cal M}$. If $\mathfrak{h}$ is the Cartan subalgebra of $\mathfrak{g}$, and the latter has rank $l$, then ${\cal U}(\mathfrak{h})$ can be identified with the subalgebra of polynomials within $C^{\infty}(\mathbb{R}^l)$. However, the restriction of the Poisson structure (\ref{pajareshijodeputa}) from ${\cal U}(\mathfrak{g})$ to ${\cal U}(\mathfrak{h})$ vanishes identically, because $\mathfrak{h}$ is abelian. We will see next that ${\cal U}(\mathfrak{g})$ can be naturally associated with a certain algebra of functions. This is best done by specifying a representation for $\mathfrak{g}$, which brings us to our next point.

Eqns. (\ref{cesargomezmekagoentuputakaramarikondemierda})--(\ref{ramayoquetepartaunrayo}) above hold for any abstract  Lie algebra $\mathfrak{g}$ (simple and compact). Given now a faithful $d$--dimensional representation for $\mathfrak{g}$, eqns. (\ref{cesargomezmekagoentuputakaramarikondemierda})--(\ref{ramayoquetepartaunrayo}) above are represented as (antisymmetrised sums of) compositions of elements of ${\rm End}\,(\mathbb{R}^d)$, and we have a (representation--dependent) isomorphism 
\begin{equation}
{\cal U}(\mathfrak{g})\simeq {\rm End}\,(\mathbb{R}^d).
\label{barbonquetepartaunrayo}
\end{equation}
{}Further fixing a basis on $\mathbb{R}^d$, endomorphisms are represented by matrices, and the associative multiplication law on ${\cal U}(\mathfrak{g})$ becomes matrix multiplication. This gives isomorphisms 
\begin{equation}
{\rm End}\,(\mathbb{R}^d)\simeq {\rm Mat}_{d\times d}(\mathbb{R})\simeq{\cal U}(\mathfrak{g}),
\label{labastidajodete}
\end{equation}
and eqns. (\ref{cesargomezmekagoentuputakaramarikondemierda})--(\ref{ramayoquetepartaunrayo}) above can be written as antisymmetrised sums of powers of $(d\times d)$ matrices. Thus elements $F\in{\cal U}(\mathfrak{g})$ are represented by $(d\times d)$--dimensional matrices with entries $F_{jm}$. The $F_{jm}$ are the coordinate functions of $F\in {\cal U}(\mathfrak{g})$ in the given representation. As such they are polynomials of arbitrary degree in the coordinates $x^1,\ldots,x^{d^2}$ on $\mathbb{R}^{d^2}$; these polynomials are homogeneous of degree 1 when $F\in \mathfrak{g}$. So, in the given representation for $\mathfrak{g}$, the matrix entries $F_{jm}$ specifying $F\in{\cal U}(\mathfrak{g})$ are polynomial functions,
\begin{equation}
F_{jm}\colon\mathbb{R}^{d^2}\longrightarrow\mathbb{R}\qquad j,m=1,\ldots, d,
\label{labastidahijoputa}
\end{equation}
and the nonabelian, pointwise multiplication law on ${\cal U}(\mathfrak{g})$ is matrix multiplication,
\begin{equation}
(FG)_{jk}=\sum_{m=1}^{d}F_{jm}G_{mk}, \qquad j,k=1,\ldots, d.
\label{labacabron}
\end{equation}

\subsection{Quantum}\label{barbonmaricon}

On the linear space $\mathbb{R}^{N}$ we have an associative, commutative  algebra of functions $C^{\infty}(\mathbb{R}^{N})$ with respect to the pointwise product: if $f,g\in C^{\infty}(\mathbb{R}^{N})$, then their pointwise product is the function 
\begin{equation}
(f\cdot g)\colon\mathbb{R}^N\longrightarrow \mathbb{R}\qquad (f\cdot g)(x):=f(x)g(x) \;\; \forall x\in\mathbb{R}.
\label{barbonkabronquetepartaunrayo}
\end{equation}
Let a Poisson structure $\left\{\cdot\,,\cdot\right\}_{\rm Poisson}$ be given on $\mathbb{R}^{N}$. Picking coordinates $x^1, \ldots, x^N$ on $\mathbb{R}^{N}$ we can write
\begin{equation}
\left\{f,g\right\}_{\rm Poisson}(x)=\Omega^{jm}(x)\partial_jf(x)\partial_mg(x),
\label{ramallokaka}
\end{equation}
where $\Omega^{jm}(x)=-\Omega^{mj}(x)$ is the matrix of $\left\{\cdot\,,\cdot\right\}_{\rm Poisson}$ at $x\in\mathbb{R}^N$, and 
all products involved are pointwise. Associated with  $\left\{\cdot\,,\cdot\right\}_{\rm Poisson}$ there is a $\star$--product \cite{KONTSEVICH} which is an associative, noncommutative deformation of the pointwise product on $C^{\infty}(\mathbb{R}^{N})$, such that
\begin{equation}
f\star g=f\cdot g + O(\hbar)
\label{labastidamecagoentuputasombramarikon}
\end{equation}
and
\begin{equation}
\left\{f,g\right\}_{\rm Poisson}=\frac{1}{{\rm i}\hbar}\left(f\star g-g\star f\right)+O(\hbar)
\label{labastidamecagoentuputacaracabron}
\end{equation}
for all $f,g\in C^{\infty}(\mathbb{R}^N)$. In fact the Kontsevich $\star$--product is uniquely determined (up to gauge equivalence) by the given Poisson structure $\left\{\cdot\,,\cdot\right\}_{\rm Poisson}$ on ${\cal M}$ \cite{KONTSEVICH}. Replacing the pointwise product on $C^{\infty}(\mathbb{R}^{N})$ with the Kontsevich $\star$--product provides a quantum deformation of this latter algebra, denoted $C^{\star}(\mathbb{R}^{N})$. We should point out that the Kontsevich $\star$--product reduces to Gr\"onewald--Moyal's \cite{GRONEWALD, MOYAL}
\begin{equation} 
(f\star g)(x)=f(x)\,\exp{\left({\rm i}\hbar\,\stackrel{\leftarrow}{\partial_j}\Omega^{jm}\stackrel{\rightarrow}{\partial_m}\right)}\,g(x)
\label{marikitasjaviermas}
\end{equation}
in the case when the Poisson structure $\Omega^{jm}$ is constant, {\it i.e.}, independent of $x\in\mathbb{R}^N$. However in our setup the Kirillov--Poisson brackets (\ref{barbonchupamelapolla}) are the natural choice. The reason is the explicit presence of the structure constants $f^{jm}_k$ of $\mathfrak{g}$ in the Kirillov--Poisson brackets: for the latter we have $\Omega^{jm}(x)=f^{jm}_kx^k$, where the $x^k$ are coordinates on $\mathfrak{g}^*$.

We define the nonabelian, noncommutative algebra ${\cal U}^{\star}(\mathfrak{g})$ as the one resulting from ${\cal U}(\mathfrak{g})$
upon replacing the matrix pointwise multiplication (\ref{labacabron}) with the matrix $\star$--product
\begin{equation}
(F\star G)_{jk}=\sum_{m=1}^{d}F_{jm}\star G_{mk} \qquad j,k=1,\ldots, d,
\label{labamaricon}
\end{equation}
where $F_{jm}\star G_{mk}$ is the Kontsevich $\star$--product of functions on $\mathbb{R}^{d^2}$. That is, we are setting $N=d^2$ in eqns. (\ref{barbonkabronquetepartaunrayo})--(\ref{labastidamecagoentuputacaracabron}) above. However this requires previous Poisson brackets $\left\{\cdot\,,\cdot\right\}_{\rm Poisson}$ on the algebra $C^{\infty}(\mathbb{R}^{d^2})$. In order to define them we observe that
\begin{equation}
\mathfrak{g}^*\simeq\mathfrak{g}\subset{\rm End}\,(\mathbb{R}^d)\simeq\mathbb{R}^{d^2},
\label{alvarezgaumecabron}
\end{equation}
where the first $\simeq$ sign is due to the Killing form, and the inclusion sign reminds us that not all endomorphisms qualify as elements of $\mathfrak{g}$. For example the identity endomorphism, having nonzero trace, cannot belong to any simple $\mathfrak{g}$. It follows that
\begin{equation}
C^{\infty}(\mathfrak{g}^*)\subset C^{\infty}(\mathbb{R}^{d^2}).
\label{alvarezgaumemecagoentuputacaramarikondemierda}
\end{equation}
By eqn. (\ref{barbonchupamelapolla}), at least a subalgebra of $C^{\infty}(\mathbb{R}^{d^2})$ supports natural Poisson brackets: the Kirillov brackets on $C^{\infty}(\mathfrak{g}^*)$. One can try and extend the latter to all of $C^{\infty}(\mathbb{R}^{d^2})$, and in fact any extension will do the job. However any such extension will be redundant since the $\star$--product will only enter our equations through antisymmetrised expressions. This is so because the conditions restricting an endomorphism $X\in{\rm End}\,(\mathbb{R}^d)$ to be also an element of the Lie algebra $\mathfrak{g}$ ({\it e.g.}, the tracelessness condition) carry over unchanged to the $\hbar$--deformed case. Hence the resulting antisymmetrised $\star$--matrix multiplication on ${\cal U}^{\star}(\mathfrak{g})$ is independent of which extension is picked for the Kirillov brackets. To summarise, eqn. (\ref{labamaricon}) correctly defines a nonabelian, noncommutative multiplication on the algebra ${\cal U}^{\star}(\mathfrak{g})$. Moreover, the Poisson structure picked to define the $\star$--product is the natural one, namely, the Kirillov brackets. Next we have the $\star$--isomorphism 
\begin{equation}
{\cal U}^{\star}(\mathfrak{g})\simeq {\rm End}^{\star}(\mathbb{R}^{d});
\label{pajareseresuntontodebaba}
\end{equation}
the superscript $\star$ reminds us that matrices are to be multiplied according to eqn. (\ref{labamaricon}). Picking a basis of vectors in $\mathbb{R}^d$, the algebra ${\rm End}^{\star}(\mathbb{R}^{d})$ is $\star$--isomorphic to the algebra of $(d\times d)$ matrices whose entries are $\star$--polynomial functions of arbitrary degree in the $x^1,\ldots,x^{d^2}$.

{}Finally the quantum dynamics on ${\cal U}^{\star}(\mathfrak{g})$ is described by $2j$--fold Poisson brackets, 
\begin{equation}
\{\cdot\,,\cdots,\cdot \}_{\omega_{2j+1}}^{\star}\colon {\cal U}^{\star}(\mathfrak{g})\times{}^{(2j)\atop\ldots}\times 
{\cal U}^{\star}(\mathfrak{g})
\longrightarrow {\cal U}^{\star}(\mathfrak{g}),
\label{pajaresmaricon}
\end{equation}
that can be obtained from the classical $2j$--fold Poisson brackets (\ref{pajareshijodeputa}) by just replacing all pointwise matrix products (\ref{labacabron}) with $\star$--matrix products, as per eqn. (\ref{labamaricon}). The result provides an $\hbar$--deformation of the classical $2j$--fold brackets (\ref{pajareshijodeputa}), to which it reduces in the limit $\hbar\to 0$. The time evolution of an observable $F\in {\cal U}^{\star}(\mathfrak{g})$
is governed by the equation
\begin{equation}
\dot F = \{F,H_2,\ldots, H_{2j}\}^{\star}_{\omega_{2j+1}},
\label{labakaka}
\end{equation}
which is the quantum analogue of the classical equation of motion (\ref{ramayoquetepartaunrayo}). The latter can be obtained from the above by letting $\hbar\to 0$.

\section{The connection with D--branes}\label{labastidaquetefollen}

Our previous correspondence between coordinates and matrices is essential in order to understand the latter as a natural generalisation of the former. While standard geometry has number--valued functions as coordinates, matrix--valued functions arise naturally as {\it transverse coordinate functions}\/ for D$p$--branes \cite{WITTEN}. Next we demonstrate that the Poisson multibrackets of section \ref{casposoramallo} are appropriate to describe the classical and quantum dynamics of the transverse coordinates to branes.

The superposition of $n$ parallel, identical D$p$--branes produces a $\mathfrak{u}(n)$ gauge theory on their common $(p+1)$--dimensional worldvolume \cite{WITTEN}. Now $\mathfrak{u}(n)=\mathfrak{u}(1)\times \mathfrak{su}(n)$ is not simple, but separating out the centre--of--mass motion we are left with the simple algebra $\mathfrak{su}(n)$. Let $A_{\mu}$ be an $\mathfrak{su}(n)$--valued gauge field on the D$p$--brane stack, and separate its components into longitudinal and transverse parts to the D$p$--branes, $A_{\mu}=(A_l, A_t)$. Longitudinal components $A_l$ are then adjoint--valued $\mathfrak{su}(n)$ matrices, {\it i.e.}, Yang--Mills gauge fields. Transverse components $A_t$ describe D$p$--brane fluctuations that are orthogonal to the D$p$--branes themselves. They are thus identified with transverse coordinates, so they are more properly denoted $X_l$ instead of $A_l$. Modulo numerical factors, the bosonic part of the mechanical action of super Yang--Mills theory dimensionally reduced to $p+1$ dimensions is \cite{SZABO}
\begin{equation}
S_{\rm YM}^{(p+1)}=\int{\rm d}^{p+1}\xi\,{\rm tr}\,({\cal F}_{ll'}^2+2{\cal F}^2_{lt}+{\cal F}^2_{tt'}),
\label{labastidachupamelapollamarikondemierda}
\end{equation}
where $l,l'$ are longitudinal indices, $t,t'$ are transverse, and the trace is taken is the adjoint representation. Dirichlet boundary conditions remove all derivatives in the $t$ directions, and (again up to numerical factors) eqn. (\ref{labastidachupamelapollamarikondemierda}) becomes
\begin{equation}
S_{\rm YM}^{(p+1)}=\int{\rm d}^{p+1}\xi\,{\rm tr}\,{\cal F}_{ll'}^2 - \int{\rm d}^{p+1}\xi\,{\rm tr}\,\left(\frac{1}{2} (D_lX^t)^2-\frac{1}{4}[X^t,X^{t'}]^2
\right),
\label{cesargomezquetefollenkabrondemierda}
\end{equation}
where $D_lX_t=\partial_lX_t+{\rm i}[A_l, X_t]$ is the longitudinal, gauge--covariant derivative of transverse coordinates. The appearance of matrix--valued coordinate functions can be motivated in the relation of D$p$--branes to Chan--Paton factors via T--duality \cite{SZABO}. For $p=-1$ (the case of instantons), all spacelike directions are transverse; for $p=0$, all but one. The latter is the important case of the M(atrix) model \cite{BFSS} of M--theory, where the limit $n\to\infty$ is taken. What follows can be regarded as applying to the lagrangian density describing the transverse coordinates $X_t$, which is the integrand of the second summand on the right--hand side of (\ref{cesargomezquetefollenkabrondemierda}); the corresponding action will be the volume integral of this lagrangian over {\it transverse}\/ space. (Notice that the integral (\ref{cesargomezquetefollenkabrondemierda}) extends over {\it longitudinal}\/ space instead). Thus our action integral reads, in the 11 dimensions of M--theory \cite{SZABO},
\begin{equation}
S_{\rm transverse}=\int{\rm d}^{10-p}\xi\,{\rm tr}\,\left( (D_lX^t)^2-\frac{1}{2}[X^t,X^{t'}]^2
\right),
\label{ramallomechupalapolla}
\end{equation}
as always up to overall factors.

We recall that $\mathfrak{su}(n)$ has $n-1$ simple roots \cite{HUMPHREYS}, 
\begin{equation}
\alpha_1={\bf e}_1-{\bf e}_2,\quad \alpha_2={\bf e}_2-{\bf e}_3,\quad \ldots\quad \alpha_{n-1}={\bf e}_{n-1}-{\bf e}_n, 
\label{javiermasmarikondeplaya}
\end{equation}
the ${\bf e}_j$, $j=1,\ldots n$, being an orthonormal basis in $\mathbb{R}^n$. The simple root $\alpha_j={\bf e}_j-{\bf e}_{j+1}$ can be understood as corresponding to a string connecting the D--branes $j$ and $j+1$ within the stack of $n$ coincident D--branes. Nonsimple, positive roots such as, {\it e.g.}, $\beta=\alpha_j+\alpha_{j+2}$, correspond to strings connecting nonadjacent D--branes; negative roots correspond to oppositely oriented strings. (The strings themselves are stretched only when the corresponding D--branes are separated, thus breaking the $\mathfrak{su}(n)$ gauge symmetry to a subalgebra \cite{WITTEN}).  Separating now the $n$--th D--brane from the remaining $n-1$ coincident D--branes reduces the gauge symmetry down to $\mathfrak{su}(n-1)\times\mathfrak{u}(1)$; this corresponds to eliminating the simple root $\alpha_{n-1}$. In this process the $\mathfrak{su}(n)$ generators $e_{\pm\alpha_{n-1}}$ are removed, but not so their diagonal commutator $h_{\alpha_{n-1}}=[e_{\alpha_{n-1}},e_{-\alpha_{n-1}}]$, which remains as the generator of the $\mathfrak{u}(1)$ corresponding to the separated brane. Further separating out more branes from the stack one can reduce this matrix dynamics all the way down to $\mathfrak{u}(1)^{\times n}$. 

In the given representation we can arrange to have $e_{\alpha_j}^{\dagger}=e_{-\alpha_j}$ for all $j=1,\ldots, n$. That is, the adjoint of the generator $e_{\alpha_j}$, with $\alpha_j$ a simple root, is the generator $e_{-\alpha_j}$ corresponding to the opposite root. Within ${\cal U}(\mathfrak{su}(n))$ let us consider the $2n-2$ selfadjoint matrices defined as
\begin{equation}
H_j^{(\pm)}:=\frac{1}{2}\sum_{l=1}^{j}\left(e_{\alpha_l}e_{-\alpha_l}\pm e_{-\alpha_l}e_{\alpha_l}\right), \qquad j=1, \ldots, n-1.
\label{ramallocasposoquetedenpurculo}
\end{equation}
We claim that the $H_j^{(\pm)}$ play the role of selfadjoint Hamiltonian operators (matrices) for the dynamics (\ref{ramallomechupalapolla}) describing the coordinates transverse to a stack of $n$ coincident D$p$--branes. In order to justify our claim we first recall that the integral (\ref{ramallomechupalapolla}) does not extend over the time coordinate, because time is longitudinal. Hence the canonical Hamiltonian that one would naively construct out of (\ref{ramallomechupalapolla}), and the corresponding Poisson brackets, are meaningless. We need a geometric, Lie--algebraic prescription to give  the dynamics (\ref{ramallomechupalapolla}) a meaning.
Let us further recall that $\mathfrak{su}(n)$ has the nontrivial cohomology cocycles $\omega_3, \omega_5$, $\ldots$, $\omega_{2n-1}$ \cite{KNAPP}. The cohomology of the corresponding Lie group, $SU(n)$, is a product of spheres, $S^3\times S^5\times \ldots\times S^{2n-1}$ \cite{GOLDBERG}. Any such sphere $S^{2j-1}$, for $j=2,\ldots, n$, is the submanifold of $\mathbb{R}^{2j}$ defined by
\begin{equation}
\sum_{l=1}^{2j}(x^l)^2=1.
\label{barbonmatrikonquetelametanporkulo}
\end{equation}
Pairwise grouping the $2j$ real coordinates $x^l$ into $j$ complex ones $z^l$ on $\mathbb{C}^{j}$, (\ref{barbonmatrikonquetelametanporkulo}) becomes
\begin{equation}
\sum_{l=1}^{j} z^l \bar z^l=1,
\label{labastidamechupalapolla}
\end{equation}
which we write more suggestively as
\begin{equation}
\frac{1}{2}\sum_{l=1}^{j}\left( z^l \bar z^l+\bar z^l z^l\right)=1.
\label{ramalloquetelametanporculo}
\end{equation}
Identifying the generator $e_{\alpha_l}$ with the complex variable $z^l$ and $e_{-\alpha_l}$ with its complex conjugate $\bar z^l$, the $n-1$ operators $H_j^{(+)}$ have a clear geometric origin. The remaining $n-1$ matrices, given by $H_j^{(-)}$, are linearly independent of the $H_j^{(+)}$. The $H_j^{(-)}$ actually equal a sum of the diagonal Cartan generators for the $\mathfrak{su(2)}$ subalgebras within $\mathfrak{su(n)}$, but this fact is immaterial to what follows. We will presently provide a physical interpretation for the appearance of 2 Hamiltonians, $H_j^{(+)}$ and $H_j^{(-)}$, for each value of $j=1, \ldots, n-1$. Altogether the $2n-2$ operators $H_j^{(\pm)}$ form a linearly independent set of selfadjoint matrices within ${\cal U}(\mathfrak{su}(n))$. 

We can write down a classical evolution equation for $\mathfrak{su}(n)$ involving all the Hamiltonians (\ref{ramallocasposoquetedenpurculo}). For this we consider the top cocycle $\omega_{2n-1}$, whose Poisson multibrackets involve $2n-2$ entries and set, for any $F\in{\cal U}(\mathfrak{su}(n))$, 
\begin{equation}
\dot F = \left\{F,  H_{1}^{(+)}, H_{1}^{(-)}, H_{2}^{(+)}, H_{2}^{(-)}\ldots,  H_{n-1}^{(+)}\right\}_{\omega_{2n-1}}.
\label{cesargomezmekagoentuputakaramarikondeplaya}
\end{equation}
We observe that the final entry above is $H_{n-1}^{(+)}$, while $H_{n-1}^{(-)}$ is missing. In fact we could just as well write 
\begin{equation}
\dot F = \left\{F,  H_{1}^{(+)}, H_{1}^{(-)}, H_{2}^{(+)}, H_{2}^{(-)}\ldots,  H_{n-1}^{(-)}\right\}_{\omega_{2n-1}}.
\label{alvarezgaumekaka}
\end{equation}
Thus we have two independent evolution equations, (\ref{cesargomezmekagoentuputakaramarikondeplaya}) and (\ref{alvarezgaumekaka}), that we can regard as corresponding to the two different orientations that the top cohomology cocycle $S^{2n-1}$ can have. This makes sense since, in the absence of a continuously flowing time parameter, there is no canonical choice of an orientation; the latter has to be determined geometrically. All other (lower--dimensional) cocycles are represented within (\ref{cesargomezmekagoentuputakaramarikondeplaya}) and (\ref{alvarezgaumekaka}) in their two possible orientations. To summarise, we have the classical equations of motion:
\begin{equation}
\dot F = \left\{F,  H_{1}^{(+)}, H_{1}^{(-)}, H_{2}^{(+)}, H_{2}^{(-)}\ldots,  H_{n-1}^{(\pm)}\right\}_{\omega_{2n-1}}.
\label{labastidacasposoquetedenpurculo}
\end{equation}
Upon quantisation, the above becomes
\begin{equation}
\dot F = \left\{F,  H_{1}^{(+)}, H_{1}^{(-)}, H_{2}^{(+)}, H_{2}^{(-)}\ldots,  H_{n-1}^{(\pm)}\right\}_{\omega_{2n-1}}^{\star}.
\label{cesargomezmekagoentuputakaramarikonazodemierda}
\end{equation}

\section{Discussion}\label{mierdaparatodos}

Apparently there is nothing compelling about the stack of $n$ coincident D--branes that forces one to describe its mechanics using Poisson multibrackets of \cite{BUENO}. The Lie algebra $\mathfrak{su}(n)$ arises naturally when superimposing $n$ D--branes, but the equations of motion we have written down have an algebraic origin in the Lie algebra cohomology, that is apparently independent of any branes whatsoever. After all one could just as well continue to use the standard binary Poisson brackets, with the quadratic Casimir of the Lie algebra as the Hamiltonian. Nothing seems to require more than one Hamiltonian, {\it i.e.}, more than one generator of translations along a timelike coordinate. 

However, the time coordinate itself is parallel to the brane, so all transverse coordinates are spacelike. In particular there is no transverse time to a D--brane. Transverse coordinates to a brane are all matrix--valued and all spacelike. The goal we set out to achieve was the description of the {\it transverse}\/ directions to a brane. So, if the Hamiltonian is the generator of time translations, with time being longitudinal, either there is no Hamiltonian at all, or there is no reason to restrict to just one Hamiltonian. 
In this article we adopt this latter point of view. This opens up many possibilities for evolution equations, now that time evolution becomes an algebraic property instead of a smooth evolution along a distinguished, continuous parameter.

The lesson we learn is that evolution equations for transverse, Lie--algebra valued coordinate functions such as those considered here are determined by the gauge symmetry present in the branes, rather than by the {\it coordinate}\/ aspect of those coordinate functions. In other words, the {\it Lie--algebra}\/ aspect prevails over the {\it coordinate}\/ aspect. This is in accord with branes as worldvolumes for (supersymmetric) gauge theories, at least at low energies \cite{SZABO}. The corresponding dynamics must therefore take this fact into account; the Poisson multibrackets considered here do precisely that. Last but not least, matrix--valued coordinate functions provide an interesting example of noncommutative geometry \cite{CONNES, LANDI}.

\vskip1cm
\noindent {\bf Acknowledgements}  J.M.I.  thanks Max--Planck--Institut f\"ur Gravitationsphysik, Albert--Einstein--Institut (Golm, Germany) for hospitality during the preparation of this article. This work has been supported by Generalitat Valenciana (Spain).

\end{document}